\documentclass[letterpaper, 10 pt,conference]{ieeeconf}  
\usepackage{amssymb}
\usepackage{amsmath,mathrsfs}
\usepackage{color}
\usepackage{graphicx}
\usepackage{mathtools}
\usepackage{lineno,hyperref}
\usepackage{cleveref}
\usepackage{bm}
\let\labelindent\relax
\usepackage{enumitem}
\usepackage{tikz}
\usepackage{comment}
\usepackage{algorithm}
\usepackage{algorithmicx}
\usepackage{adjustbox}
\usepackage{algpseudocode}
\modulolinenumbers[5]
\newtheorem{theorem}{Theorem}[section]
\newtheorem{remark}{Remark}[section]

\newtheorem{lemma}[theorem]{Lemma}

\usepackage{tikz}
\newcommand{\tikzline}[1]{(\protect\tikz[baseline=-0.6ex,x=1pt,y=1pt]{ \protect\draw[#1] [-] (0,0) -- (10,0);})}

\newcommand{\tikzdashdottedline}[1]{(\protect\tikz[baseline=-0.6ex,x=1pt,y=1pt]{ \protect\draw[#1,dash dot] [-] (0,0) -- (10,0);})}

\definecolor{MatlabBlue}{rgb}    {0     , 0.4470, 0.7410}
\definecolor{MatlabRed}{rgb}     {0.8500, 0.3250, 0.0980}
\definecolor{MatlabYellow}{rgb}  {0.9290, 0.6940, 0.1250}
\definecolor{MatlabPurple}{rgb}  {0.4940, 0.1840, 0.5560}
\definecolor{MatlabGreen}{rgb}   {0.4660, 0.6740, 0.1880}
\definecolor{MatlabBabyBlue}{rgb}{0.3010, 0.7450, 0.9330}
\definecolor{MatlabGray}{rgb}{0.5, 0.5, 0.5}
\definecolor{MatlabLightGray}{rgb}{0.75, 0.75, 0.75}
\definecolor{MatlabBlack}{rgb}{0, 0, 0}
\definecolor{MatlabLightGray4}{rgb}{0.875, 0.875, 0.875}
\definecolor{MatlabLightGray3}{rgb}{0.85, 0.85, 0.85}
\definecolor{MatlabLightGray2}{rgb}{0.775, 0.775, 0.775}
\definecolor{MatlabLightGray1}{rgb}{0.7, 0.7, 0.7}
\definecolor{MatlabGray20}{rgb}{0.2, 0.2, 0.2}
\definecolor{MatlabGray30}{rgb}{0.3, 0.3, 0.3}
\definecolor{MatlabGray40}{rgb}{0.4, 0.4, 0.4}
\definecolor{MatlabGray50}{rgb}{0.5, 0.5, 0.5}
\definecolor{MatlabGray60}{rgb}{0.6, 0.6, 0.6}
\definecolor{MatlabGray70}{rgb}{0.7, 0.7, 0.7}
\definecolor{MatlabGray80}{rgb}{0.8, 0.8, 0.8}
\definecolor{MatlabGray85}{rgb}{0.85, 0.85, 0.85}
\definecolor{MatlabGray90}{rgb}{0.9, 0.9, 0.9}

\usepackage{multirow}
\usepackage{color}
\usepackage{cite}
\usepackage{mathtools}

\providecommand{\keywords}[1]{\textbf{\textit{Index terms---}}}

\IEEEoverridecommandlockouts                              
\title{\LARGE \bf Identification of additive multivariable continuous-time systems}

\author{Maarten van der Hulst$^{1,*}$, Rodrigo A. Gonz\'alez$^1$, Koen Classens$^1$, Nic Dirkx$^2$, \\ Jeroen van de Wijdeven$^2$, Tom Oomen$^{1,3}$ 
\thanks{ ${ }^1$ Dept. of Mechanical Engineering, Eindhoven University of Technology, The Netherlands. ${ }^2$ ASML, Veldhoven, The Netherlands. ${ }^3$Delft Center for Systems and Control, Delft University of Technology, The Netherlands. This project is funded by Holland High Tech | TKI HSTM via the PPP Innovation Scheme (PPP-I) for public-private partnerships. ${ }^*$Corresponding author, Email: \tt M.v.d.hulst@tue.nl.
}%
}

\usepackage{eso-pic}
\AddToShipoutPictureBG*{%
\AtPageUpperLeft{%
\setlength\unitlength{1in}%
\hspace*{\dimexpr0.5\paperwidth\relax}
\makebox(0,-1.75)[c]{
\begin{tabular}{c c}
Maarten van der Hulst \emph{et al.},
Identification of additive multivariable continuous-time systems, \\
accepted to LCSS, uploaded to ArXiv  \today \\
\end{tabular}}}}

\begin{document}
\title{Identification of additive multivariable continuous-time systems}

\author{Maarten van der Hulst$^{1,*}$, Rodrigo A. Gonz\'alez$^1$, Koen Classens$^1$, Nic Dirkx$^2$, \\ Jeroen van de Wijdeven$^2$, Tom Oomen$^{1,3}$ 
\thanks{ ${ }^1$ Dept. of Mechanical Engineering, Eindhoven University of Technology, The Netherlands. ${ }^2$ ASML, Veldhoven, The Netherlands. ${ }^3$Delft Center for Systems and Control, Delft University of Technology, The Netherlands. This project is funded by Holland High Tech | TKI HSTM via the PPP Innovation Scheme (PPP-I) for public-private partnerships. ${ }^*$Corresponding author, Email: \tt M.v.d.hulst@tue.nl.
}%
}

\maketitle
\thispagestyle{empty}


\begin{abstract}
Multivariable parametric models are critical for designing, controlling, and optimizing the performance of engineered systems. The main aim of this paper is to develop a parametric identification
strategy that delivers accurate and physically relevant models of multivariable systems using time-domain data. The introduced approach adopts an additive model structure, providing a parsimonious and interpretable representation of many physical systems, and applies a refined instrumental variable-based estimation algorithm. The developed identification method enables the estimation of multivariable parametric additive models in continuous time and is applicable to both open- and closed-loop systems. The performance of the estimator is demonstrated through numerical simulations and experimentally validated on a flexible beam system.
\end{abstract}
\section{Introduction}
System identification is critical in modeling and analyzing the dynamic behavior of multivariable systems in many engineering domains. The obtained data-based models are important for analyzing and predicting system dynamics, as well as for designing high-performance observers and control strategies. Many engineered systems contain a large number of inputs and outputs, and the ability of identification methods to effectively handle their inherent multivariable nature is vital to further optimize system performance.

Traditional linear system identification approaches for multivariable systems often include model structures that do not exploit the structure of the considered physical system \cite{Soderstrom2001SystemIdentification}. Examples include unfactored common denominator models, expressed as the quotient of a numerator matrix polynomial and scalar denominator polynomial, or matrix fractional descriptions (MFDs). The literature on these model parameterizations is extensive \cite{Glover1974ParametrizationsIdentifiability, Correa1984Pseudo-canonicalModels, Vayssettes2016NewApproach},  yet MFDs may not provide the most parsimonious or physically relevant model descriptions when considering practical applications in, e.g., vibrational analysis and flexible motion systems \cite{Gawronski2004AdvancedStructures, Oomen2018AdvancedSystems}. These systems are often modeled in a modal form, consisting of a sum of low-order transfer functions, where each term in the additive decomposition corresponds to a specific resonant mode of the system. Additive models are related to unfactored transfer functions through partial fraction expansion and are encountered in many engineering fields, such as RLC circuit modeling \cite{Lange2021BroadbandShape} and thermal analysis of machine frames \cite{Zhu2008RobustConcept}. Additive model structures offer several advantages, such as providing more physically insightful models for model updating \cite{DaSilva2008DesignOptimization}, and online monitoring techniques \cite{Classens2022FaultResonances}, while also providing more parsimonious parameterizations leading to improved statistical estimation properties \cite{Soderstrom2001SystemIdentification}. Furthermore, these structures contribute to numerically more stable algorithms \cite{Pintelon2012SystemIdentification} and improve numerical conditioning, particularly for parametric identification of high-order systems \cite{Gilson2018AIdentification}.

Accurate modeling of physical systems benefits from approaches that align with the underlying dynamics. In this regard, the estimation of continuous-time models \cite{Garnier2008IdentificationData} offers advantages over discrete-time models as they facilitate the direct incorporation of \textit{a priori} knowledge, such as the relative degree of the system, which for discrete-time models is more challenging due to the presence of sampling zeros \cite{Rao2006IdentificationSystems}. Moreover, direct continuous-time identification is well suited for fast or non-uniformly sampled settings, and the parameters directly correspond to physical quantities, aiding model interpretability \cite{Garnier2015DirectApplications}. 

The estimation of additive model parameterizations has mainly been considered for the single-input single-output (SISO) setting. Examples include vector fitting procedures \cite{Semlyen1999RationalFitting}, which estimates first-order pole models, as well as the identification of nonlinear finite impulse responses and generalized Hammerstein models \cite{Bai2005IdentificationSystems, Bai2008IdentificationModels}. Recent results in additive system identification include the direct continuous-time identification method introduced in \cite{Gonzalez2024IdentificationClosed-loop}, based on the simplified refined instrumental variable method (SRIVC) \cite{Young1980RefinedAnalysis}, as well as block coordinate descent techniques \cite{ClassensRecursiveSystems,Gonzalez2023ParsimoniousApproach}. In contrast to these SISO methods, many multi-input multi-output (MIMO) identification strategies have been developed that consider non-additive structures. Estimating MIMO models is inherently more challenging due to increased model complexity and typically larger-dimensional estimation problems. Most of these methods are categorizable in pseudo-linear regression-based approaches \cite{Blom2010MultivariableRegression, Akroum2009ExtendingModels, Soderstrom2012AProblems}, gradient descent methods \cite{Bayard1994High-orderResults} and subspace techniques \cite{Ohsumi2002SubspaceApproach}. 

Although important developments in additive identification have been made, existing techniques are either limited to SISO systems, do not explicitly address the closed-loop setting or are not directly applicable to time-domain identification of continuous-time models. This paper presents a comprehensive time-domain identification method for estimating additive MIMO systems in continuous time applicable to both open- and closed-loop configurations. The proposed method extends the work in \cite{Gonzalez2024IdentificationClosed-loop} and can be viewed as a time-domain counterpart to the frequency-domain approach introduced in \cite{vanderHulst2025FrequencyStage}. The main contributions of this paper are:

\begin{enumerate}[label=C\arabic*]
    \item A time-domain identification strategy for estimating continuous-time models of MIMO systems in additive transfer function form, applicable to both open-loop and closed-loop scenarios.
    
    \item Experimental validation of the developed identification method by application to a MIMO flexible-beam system. 
\end{enumerate}

This paper is structured as follows. Section 2 introduces the additive model structure and presents the identification problem considered. In Section 3, the developed identification strategy is derived. Numerical simulations are provided in Section 4, with experimental validation of the method in Section 5. Finally, conclusions are given in Section 6.

\textit{Notation:} Scalars, vectors and matrices are written as $x$, $\mathbf{x}$ and $\mathbf{X}$, respectively. If $ \mathbf{x} \in \mathbb{R}^n $ is a column vector and $ \mathbf{Q} \in \mathbb{C}^{n \times n} $ is a Hermitian matrix, then the weighted 2-norm is given by $ \|\mathbf{x}\|_{\mathbf{Q}} = \sqrt{\mathbf{x}^{\top} \mathbf{Q} \mathbf{x}} $. The Kronecker product is represented by the $ \otimes $ operator and for $ \mathbf{X} = [\mathbf{x}_1, \ldots, \mathbf{x}_n] $ with $ \mathbf{x}_i \in \mathbb{C}^n $ the operation $ \operatorname{vec}(\mathbf{X}) = [\mathbf{x}^\top_1, \ldots, \mathbf{x}^\top_n]^\top $ restructures the matrix into a vector by stacking its columns. The Heaviside operator is denoted by $p \mathbf{u}(t)=\frac{\mathrm{d}}{\mathrm{d} t}\mathbf{u}(t)$.

\section{Problem formulation}
\begin{figure}
    \centering
    \vspace{0.5em}
    \includegraphics[width=0.925\linewidth]{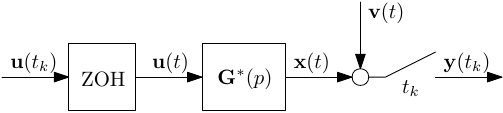}
    \includegraphics[width=0.925\linewidth]{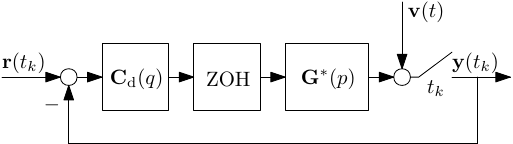}
    \caption{System acting in an open-loop setting (upper) and system acting in a closed-loop setting (lower).}
    \vspace{-0.9em}
    \label{fig: block diagram}
\end{figure}

In this section, the additive model structure is formally introduced and the identification problem is presented. Consider the causal linear and time-invariant (LTI) continuous-time MIMO system with $n_{\mathrm{u}}$ inputs and $n_{\mathrm{y}}$ outputs in additive form containing $K$ submodels
\begin{equation} \label{eq: CT - plant model}
\mathbf{x}(t) = \mathbf{G}(p,\boldsymbol{\theta})\mathbf{u}(t) = \sum_{i=1}^K\mathbf{G}_i(p,\boldsymbol{\theta})\mathbf{u}(t),
\end{equation}
where the input signal is denoted by $\mathbf{u}(t) \in \mathbb{R}^{n_{\mathrm{u}}}$ and  $\mathbf{x}(t) \in \mathbb{R}^{n_{\mathrm{y}}}$ is the unobserved noise-free plant output. The selection of $K$ can be guided by physical knowledge of the system, where, for mechanical systems, $K$ corresponds to the number of modes of the system \cite{Gawronski2004AdvancedStructures}. Each submodel $\mathbf{G}_i(p)$ is parametrized as
\begin{equation}
\mathbf{G}_i(p,\boldsymbol{\theta}) = \frac{1}{ p^{\ell_i} A_i(p)}\mathbf{B}_i(p),
\end{equation} 
where at most one submodel may include $\ell_i>0$ poles at the origin. The scalar denominator polynomial $A_i(p)$ and the matrix numerator polynomial $\mathbf{B}_i(p)$ are such that no complex number $z$ simultaneously satisfies $A_i(z) = 0$ and $ \mathbf{B}_i(z) = \mathbf{0}$. To ensure a unique characterization of $ \left\{\mathbf{G}_i(p)\right\}_{i=1}^K $, it is assumed that at most one submodel $ \mathbf{G}_i(p) $ is biproper.  The $A_i(p)$ and $\mathbf{B}_i(p)$ polynomials are parametrized as
\begin{align} 
A_i(p) &= 1 + a_{i,1} p + \ldots + a_{i,n_i}p^{n_i}, \label{eq: CT - plant polynomial A}\\
\mathbf{B}_i(p) &= \mathbf{B}_{i,0} + \mathbf{B}_{i,1} p + \ldots + \mathbf{B}_{i,m_i}p^{m_i}, \label{eq: CT - plant polynomial B}
\end{align}
where the $A_i(p)$ polynomials are stable, i.e., all roots lie in the left-half plane, and they not share any common roots. The polynomials $A_i(p)$ and $\mathbf{B}_i(p)$ are jointly described by
\begin{align} \label{eq: CT - parameter vector}
\boldsymbol{\beta} = \left[ \boldsymbol{\theta}^{\top}_1,\ \dots,\ \boldsymbol{\theta}^{\top}_K \right]^{\top},
\end{align}
with each submodel described by the vector
\begin{align} \label{eq: CT - parameter vector submodel}
\hspace{-0.75em} \boldsymbol{\theta}_i = \left[ a_{i,1}, \dots, a_{i,n_i}, \operatorname{vec}\left(\mathbf{B}_{i,0}\right)^{\top}, \dots, \operatorname{vec}\left(\mathbf{B}_{i,m_i}\right)^{\top}\right]^{\top} \hspace{-0.5em}.
\end{align}
A noisy measurement of the output is retrieved at every time instant $t=t_k,\ k=1, \ldots, N$, where $\left\{t_k\right\}_{k=1}^N$ are equidistant in time. 
That is,
\begin{equation} \label{eq: CT - output observation}
\mathbf{y}\left(t_k\right) 
= \mathbf{x}\left(t_k\right) + \mathbf{v}\left(t_k\right),
\end{equation}
where $\mathbf{v}\left(t_k\right) \in \mathbb{R}^{n_{\mathrm{y}}}$ is described by a zero-mean discrete-time random process of unknown covariance $\mathbf{\Sigma}^*$, and $\mathbf{x}\left(t_k\right)$ the sampled unobserved noise-free plant output. Two configurations are considered. In the first, the true plant $\mathbf{G}^*(p)$ operates in an open-loop setup, while in the second, the plant operates in a closed-loop configuration. The block diagrams for both the open- and closed-loop settings are shown in Figure \ref{fig: block diagram}. In the open-loop case, the output noise $\mathbf{v}(t_k)$ is assumed to be statistically independent of the input $\mathbf{u}(t_k)$, while in the closed-loop scenario, the noise is assumed to be statistically independent of the external reference signal $\mathbf{r}(t_k)\in \mathbb{R}^{n_{\mathrm{y}}}$. 
The assumed intersample behavior of the input $\mathbf{u}(t_k)$ is a zero-order hold (ZOH) and is for the closed-loop setup generated by the discrete-time LTI controller $\mathbf{C}_{\mathrm{d}}(q)$, where $q$ represents the forward-shift operator, according to the control law
\begin{equation} 
\mathbf{u}(t_k) = \mathbf{C}_{\mathrm{d}}(q) \left( \mathbf{r}(t_k) - \mathbf{y}(t_k) \right),
\end{equation}
which leads to the closed-loop expression
\begin{equation} \label{eq: CT - feedback interconnection}
\mathbf{u}\left(t_k\right) = \mathbf{S}^*_{uo}(q)\bigl(\mathbf{r}(t_k) - \mathbf{v}(t_k)\bigr),
\end{equation} 
with the control sensitivity defined as $\mathbf{S}^*_{uo}(q) = \mathbf{C}_{\mathrm{d}}(q)[{\mathbf{I}_{n_{\mathrm{y}}} + \mathbf{G}^*_{\mathrm{d}}(q)\mathbf{C}_{\mathrm{d}}(q)}]^{-1}$ and where $\mathbf{G}^*_{\mathrm{d}}(q)$ represents the ZOH discrete-time equivalent of the true plant $\mathbf{G}^*(p)$. The problem considered in this paper is to estimate the parameter vector $\boldsymbol{\beta}$ describing the additive MIMO model structure, using a dataset of input/output measurements. To this end, it is assumed that the data set $\{\mathbf{u}(t_k), \mathbf{y}(t_k)\}_{k=1}^N$ is available for the open-loop setting and the dataset $\{\mathbf{r}(t_k),\mathbf{u}(t_k), \mathbf{y}(t_k)\}_{k=1}^N$ for the closed-loop setting. In addition, the controller is assumed to be known for the closed-loop setting.

\newpage
\section{Identification strategy for continuous-time additive MIMO systems}
This section presents the identification strategy for estimating additive MIMO models and constitutes contribution~C1. 

\subsection{Identification criteria}
The identification problem is formulated using the residual output signal, which is computed as the difference between the measured system output $\mathbf{y}(t_k)$ and the simulated model output 
\begin{equation} \label{eq: CT - residual function}
\begin{aligned}
\boldsymbol{\varepsilon}(t_k,\boldsymbol{\beta}) & =\mathbf{y}(t_k) -  \sum_{i=1}^K\mathbf{G}_i(p,\boldsymbol{\theta}_i)\mathbf{u}(t_k). 
\end{aligned}
\end{equation}
\begin{remark}
In \eqref{eq: CT - residual function} a mixed notation of continuous-time filters with sampled signals is introduced. If $\mathbf{G}(p)$ is a causal continuous-time filter and $\mathbf{u}(t_k)$ is a sampled signal, then $\mathbf{G}(p)\mathbf{u}(t_k)$ implies that the signal is first interpolated using a specified intersample behavior before being filtered by $\mathbf{G}(p)$. The continuous-time output is then sampled at $t=t_k$. The notation $\mathbf{u}^\top (t_k)\mathbf{H}(p)$ can be interpreted similarly.
\end{remark}

To identify systems acting in either an open or closed-loop configuration, a correlation criterion is adopted \cite{Soderstrom2001SystemIdentification}. The parameter vector estimate $\hat{\boldsymbol{\beta}}$ is obtained as the solution to 
\begin{equation} \label{eq: CT - correlataion condition}
    \boldsymbol{0} =  \frac{1}{N}\sum^N_{k=1} \hat{\mathbf{\Phi}}\left(t_k, \boldsymbol{\beta}\right) \mathbf{\Sigma}^{-1}\boldsymbol{\varepsilon}(t_k,\boldsymbol{\beta}),
\end{equation}
with $\mathbf{\Sigma}\in \mathbb{R}^{n_\mathrm{y}\times n_\mathrm{y}}$ a positive-definite weighting matrix and $\hat{\mathbf{\Phi}}$ a matrix of instrument signals which is uncorrelated with the noise signal $\mathbf{v}(t_k)$. For the considered additive structure, this instrument matrix is partitioned as follows
\begin{equation} \label{eq: CT -  instrument matrix}
\hspace{-0.5em}    \hat{\mathbf{\Phi}}\left(t_k, \boldsymbol{\beta}\right) = \left[ \begin{array}{ccc} \hat{\mathbf{\Phi}}_{1}^\top\left(t_k, \boldsymbol{\theta}_1\right), & \ldots, & \hat{\mathbf{\Phi}}_{K}^\top\left(t_k, \boldsymbol{\theta}_K\right)\end{array}\right]^\top, 
\end{equation}
where the expression for $ \hat{\mathbf{\Phi}}_{i}$ for $i = 1,\ldots,K$ depends on the setting. For the open-loop case, the submodel instrument is selected as
\begin{align} \label{eq: CT -  OL instrument matrix submodel}
\hat{\mathbf{\Phi}}_{i}\left(t_k, \boldsymbol{\theta}_i\right) =&\ \left[ \dfrac{-p\mathbf{B}_i(p)}{p^{\ell_i} A_i^2(p)}\mathbf{u}(t_k), \dots,\ \dfrac{-p^{n_i}\mathbf{B}_i(p)}{p^{\ell_i} A_i^2(p)}\mathbf{u}(t_k), \right.\notag\\
& \hspace{-2em} \left. \dfrac{1}{p^{\ell_i} A_i(p)}\mathbf{U}^{\top}(t_k),\ \dots, \dfrac{p^{m_i}}{p^{\ell_i} A_i(p)}\mathbf{U}^{\top}(t_k)  \right]^{\top},
\end{align}
where $\mathbf{U}(t_k) = \mathbf{u}(t_k) \otimes \mathbf{I}_{n_{\mathrm{y}}}$. Under this choice of instrument, \eqref{eq: CT - correlataion condition} is interpreted as the first-order optimality condition of a least-squares minimization criterion \cite{Soderstrom2001SystemIdentification}. For the closed-loop case, the instrument corresponds to
\begin{align} \label{eq: CT -  CL instrument matrix submodel}
\hat{\mathbf{\Phi}}_{i}\left(t_k, \boldsymbol{\theta}_i\right) =&\ \left[ \dfrac{-p\mathbf{B}_i(p)}{p^{\ell_i} A_i^2(p)}\tilde{\mathbf{r}}(t_k), \dots,\ \dfrac{-p^{n_i}\mathbf{B}_i(p)}{p^{\ell_i} A_i^2(p)}\tilde{\mathbf{r}}(t_k), \right.\notag\\
& \hspace{-2em} \left. \dfrac{1}{p^{\ell_i} A_i(p)}\tilde{\mathbf{R}}^{\top}(t_k),\ \dots, \dfrac{p^{m_i}}{p^{\ell_i} A_i(p)}\tilde{\mathbf{R}}^{\top}(t_k)  \right]^{\top},
\end{align}
with $\tilde{\mathbf{r}}(t_k) = \mathbf{S}_{uo}(q)\mathbf{r}(t_k)$ the noise-free input component and $\tilde{\mathbf{R}}(t_k) = \tilde{\mathbf{r}}(t_k)\otimes \mathbf{I}_{n_{\mathrm{y}}}$. This instrument is motivated by the fact that among all instruments that produce a consistent estimator, this instrument minimizes the asymptotic covariance, which was established for the SISO case in \cite{Gonzalez2024IdentificationClosed-loop}. 

Note, that the open-loop instrument \eqref{eq: CT -  OL instrument matrix submodel} is retrieved as a special case of \eqref{eq: CT -  CL instrument matrix submodel} when the feedback interconnection is omitted. In the following subsection, a refined instrumental variable based method is introduced to estimate solutions  $\hat{\boldsymbol{\beta}}$ to the correlation condition \eqref{eq: CT - correlataion condition}.

\subsection{Refined instrumental variables for additive systems}
The correlation condition \eqref{eq: CT - correlataion condition} is non-linear in the parameter vector $\boldsymbol{\beta}$. A solution is obtained by reformulating \eqref{eq: CT - residual function} to a pseudolinear form which enables the refined instrumental variable approach. For each submodule in the additive model structure, the residual can be  reformulated into an unique pseudolinear representation, as stated in the following lemma. 
\begin{lemma} The pseudolinear regression form of the residual signal \eqref{eq: CT - residual function} corresponding to the $i$th submodel is expressed as
\begin{equation} \label{eq: CT - pseudolinear regression form i}
     \boldsymbol{\varepsilon}\left(t_k, \boldsymbol{\beta}\right) =   \tilde{\mathbf{y}}_{f,i}\left(t_k, \boldsymbol{\beta}\right) -\boldsymbol{\Phi}_i^{\top}\left(t_k, \boldsymbol{\beta}\right) \boldsymbol{\theta}_i,
\end{equation}
with the regressor matrix
\begin{align} \label{eq: CT - regressor matrix submodel}
\mathbf{\Phi}_{i}\left(t_k, \boldsymbol{\theta}_i\right) =&\ \left[ \dfrac{-p}{A_i(p)}\tilde{\mathbf{y}}_i(t_k),\dots,\ \dfrac{-p^{n_i}}{A_i(p)}\tilde{\mathbf{y}}_i(t_k) \right. \notag \\
&\left. \dfrac{1}{A_i(p)}\mathbf{U}^{\top}(t_k),\dots,\ \dfrac{p^{m_i}}{A_i(p)}\mathbf{U}^{\top}(t_k)  \right]^{\top},
\end{align}
and with $\tilde{\mathbf{y}}_{f,i}\left(t_k, \boldsymbol{\beta}\right) = A^{-1}_i(p)\tilde{\mathbf{y}}_{i}(t_k, \boldsymbol{\beta})$ the filtered version of the residual output signal
\begin{equation} \label{eq: CT - residual output submodel}
\tilde{\mathbf{y}}_{i}\left(t_k, \boldsymbol{\beta}\right) = \mathbf{y}(t_k) - \sum_{\substack{j=1, \ldots, K \\ j \neq i}} \frac{1}{p^{\ell_i} A_j(p)}\mathbf{B}_j(p)\mathbf{u}(t_k).
\end{equation}
\end{lemma} 
\vspace{0.3em}
\begin{proof}
The residual \eqref{eq: CT - residual function} is rewritten for the $i$th submodel, with $i = 1,\ldots,K$, according to
\begin{align}
\boldsymbol{\varepsilon}(t_k) & = \tilde{\mathbf{y}}_{i}\left(t_k, \boldsymbol{\beta}\right) -  \dfrac{1}{p^{\ell_i} A_i(p)}\mathbf{B}_i(p)\mathbf{u}(t_k), \\
 & = \dfrac{1}{p^{\ell_i} A_i(p)}\Bigl(p^{\ell_i} A_i(p)\tilde{\mathbf{y}}(t_k) -  \mathbf{B}_i(p)\mathbf{u}(t_k)\Bigr), \label{eq: CT - ref 1}
\end{align}
with $\tilde{\mathbf{y}}_{i}(t_k, \boldsymbol{\beta})$ defined in \eqref{eq: CT - residual output submodel}. By applying the following identity for arbitrary matrices of matching dimensions
\begin{align} \label{eq: CT - identity main}
\mathbf{X}\mathbf{b} = \left(\mathbf{b}^\top  \otimes \mathbf{I}\right) \operatorname{vec}(\mathbf{X}), 
\end{align}
and substituting \eqref{eq: CT - plant polynomial A} and \eqref{eq: CT - plant polynomial B}, each subproblem in \eqref{eq: CT - ref 1} becomes
\begin{align}
&\boldsymbol{\varepsilon}\left(t_k\right) = \dfrac{1}{A_i(p)}\tilde{\mathbf{y}}_{i}\left(t_k, \boldsymbol{\beta}\right) + \ldots + \dfrac{a_{i,n_i}p^{n_i}}{A_i(p)}\tilde{\mathbf{y}}_{i}\left(t_k, \boldsymbol{\beta}\right)  - \notag\\
&  \mathbf{U}^\top(t_k)\dfrac{\operatorname{vec}\left(\mathbf{B}_{i,0}\right)}{p^{\ell_i} A_i(p)} - \ldots - \mathbf{U}^\top(t_k)\dfrac{p^{m_i}\operatorname{vec}\left(\mathbf{B}_{i,m_i}\right)}{p^{\ell_i} A_i(p)}.
\end{align}
This expression can directly be written in the form \eqref{eq: CT - pseudolinear regression form i} by considering \eqref{eq: CT - parameter vector submodel}, thus completing the proof. 
\end{proof}
The form \eqref{eq: CT - pseudolinear regression form i} enables the correlation condition to be solved using refined instrumental variable iterations \cite{Young1980RefinedAnalysis}. Introducing the stacked signals
\begin{align}
    \mathbf{\Upsilon}\left(t_k, \boldsymbol{\beta}\right) &= \Bigl[\tilde{\mathbf{y}}_{f, 1}\left(t_k, \boldsymbol{\beta}\right),\ \ldots, \ \tilde{\mathbf{y}}_{f, K}\left(t_k, \boldsymbol{\beta}\right)\Bigr]^\top, \label{eq: CT - residual output}\\ 
    \mathbf{\Phi}\left(t_k, \boldsymbol{\beta}\right) &= \Bigl[\mathbf{\Phi}^\top_{1}\left(t_k, \boldsymbol{\theta}_1\right),\ \ldots, \ \mathbf{\Phi}^\top_{K}\left(t_k, \boldsymbol{\theta}_K\right)\Bigr]^\top, \label{eq: CT - regressor}
\end{align}
and the parameter matrix
\begin{equation} \label{eq: CT - parameter matrix}
\mathcal{B}=\left[\begin{array}{ccc}
\boldsymbol{\theta}_1 & & \mathbf{0} \\
& \ddots & \\
\mathbf{0} & & \boldsymbol{\theta}_K
\end{array}\right],
\end{equation}
which contains the elements of $\boldsymbol{\beta}$ along the block diagonal, allows the correlation condition \eqref{eq: CT - correlataion condition} for the $K$ pseudolinear regressions \eqref{eq: CT - pseudolinear regression form i} to be written according to
\begin{equation} \label{eq: CT - instrumental variable equation - matrix}
\begin{aligned}
    \sum_{k=1}^N \hat{\mathbf{\Phi}}(t_k, \boldsymbol{\beta})\mathbf{\Sigma}^{-1} \Bigl( \mathbf{\Upsilon}^\top(t_k, \boldsymbol{\beta}) - \mathbf{\Phi}^\top(t_k,\boldsymbol{\beta})\mathcal{B} \Bigr) = \mathbf{0}.
\end{aligned}
\end{equation}
The solution is obtained iteratively by fixing $\boldsymbol{\beta} = \boldsymbol{\beta}^{\langle j \rangle}$ at the $j$th iteration in (\ref{eq: CT - residual output}), the regressor (\ref{eq: CT - regressor}), and instrument matrix \eqref{eq: CT -  instrument matrix}. In addition, for minimum asymptotic covariance we require a weighting equal to the true noise covariance, i.e., $\boldsymbol{\Sigma} = \boldsymbol{\Sigma}^*$. However, since $\boldsymbol{\Sigma}^*$ is typically unknown \textit{a priori} it is estimated using 
\begin{equation} \label{eq: CT - ML noise covariance}
    \hat{\mathbf{\Sigma}}(\boldsymbol{\beta}^{\langle j \rangle}) = \frac{1}{N} \sum_{k=1}^N \boldsymbol{\varepsilon}(t_k,\boldsymbol{\beta}^{\langle j \rangle})\boldsymbol{\varepsilon}^{\top}(t_k,\boldsymbol{\beta}^{\langle j \rangle}),
\end{equation}
providing a sample maximum likelihood estimate of the true noise covariance \cite{Pintelon2012SystemIdentification}. Then, by solving for $\mathcal{B}$ in \eqref{eq: CT - instrumental variable equation - matrix}, the estimator for additive systems acting in either open or closed-loop is obtained as 
\begin{align} \label{eq: CT - estimator}
\hat{\mathcal{B}}^{\langle j+1 \rangle} &= {\left[ \sum_{k=1}^N \hat{\mathbf{\Phi}}(t_k, \boldsymbol{\beta}^{\langle j \rangle}) \hat{\mathbf{\Sigma}}^{-1}(\boldsymbol{\beta}^{\langle j \rangle})\mathbf{\Phi}^{\top}(t_k, \boldsymbol{\beta}^{\langle j \rangle})\right]^{-1} }  \times  \notag \\ 
& \hspace{2em} \sum_{k=1}^N \hat{\mathbf{\Phi}}(t_k, \boldsymbol{\beta}^{\langle j \rangle}) \hat{\mathbf{\Sigma}}^{-1}(\boldsymbol{\beta}^{\langle j \rangle}) \mathbf{\Upsilon}^{\top}(t_k, \boldsymbol{\beta}^{\langle j \rangle}),
\end{align}
where the next iteration $\boldsymbol{\beta}^{j+1}$ is extracted from the block diagonal coefficients of $\mathcal{B}^{j+1}$ described by (\ref{eq: CT - parameter matrix}). The convergence point of the iterations described by the estimator \eqref{eq: CT - estimator} provides a solution that under mild assumptions directly satisfies the correlation condition \eqref{eq: CT - correlataion condition}.

\begin{remark} Note that the iterations described by \eqref{eq: CT - estimator} recover the approach presented in \cite{Gonzalez2024IdentificationClosed-loop} when the model is SISO. In addition, for non-additive models, that is $K=1$, the SRIVC iterations are obtained in the open-loop case \cite{Young1980RefinedAnalysis}, while they correspond to CLSRIVC iterations in the closed-loop case \cite{Gilson2009RefinedIdentification}.
\end{remark}
\begin{remark}
     The proposed method is readily extendable to noise spectrum estimation by adapting the estimator for the noise covariance \eqref{eq: CT - ML noise covariance} to discrete-time ARMA model estimation, as commonly done in RIVC methods \cite{Garnier2008IdentificationData}.
\end{remark}

\subsection{Initialization}
The iterations in \eqref{eq: CT - estimator} require an initial estimate $\boldsymbol{\beta}^{\langle 0 \rangle}$ of the model parameters. This section presents a method for computing the parameters of the numerator matrix \eqref{eq: CT - plant polynomial B}, assuming fixed denominator parameters. This reduces the initialization problem to determining the initial pole locations, which are often effectively derived from nonparametric models. To this end, assume that the denominator polynomials are fixed at $\bar{A}_i(p)$, and let $\boldsymbol{\eta}$ represent the parameter vector from \eqref{eq: CT - parameter vector} without the denominator coefficients. This vector appears linearly in the residual function (\ref{eq: CT - residual function}) after applying identity \eqref{eq: CT - identity main} and therefore allows a linear regression form. The estimate  $\hat{\boldsymbol{\eta}}$ is found as the solution to the convex problem
\begin{equation} \label{eq: CT - initialization}
    \hat{\boldsymbol{\eta}} = \underset{\boldsymbol{\eta}}{\arg \min }\frac{1}{N}\sum^N_{k=1} \left\| \mathbf{y}\left(t_k\right) - \mathbf{\Phi}^{\top}\left(t_k\right) \boldsymbol{\eta} \right\|_{2}^2,
\end{equation}
where the regressor matrix $\mathbf{\Phi}$ is obtained by stacking for each submodel
\begin{align} 
\hspace{-0.4em} \mathbf{\Phi}_{i}\left(t_k\right) =&\ \left[ \dfrac{1}{p^{\ell_i}\bar{A}_i(p)}\mathbf{U}^{\top}(t_k),\dots, \dfrac{p^{m_i}}{p^{\ell_i}\bar{A}_i(p)}\mathbf{U}^{\top}(t_k) \right]^{\top}\hspace{-0.65em},
\end{align}
in a similar way as \eqref{eq: CT - regressor}. An initial estimate $\boldsymbol{\beta}^{\langle 0 \rangle}$ is thus determined by first providing initial pole locations, which enable the computation of the numerator parameters by solving \eqref{eq: CT - initialization} given data. 

\section{Simulation}

\begin{figure}[b]
\centering
    \includegraphics[width = 0.7\linewidth]{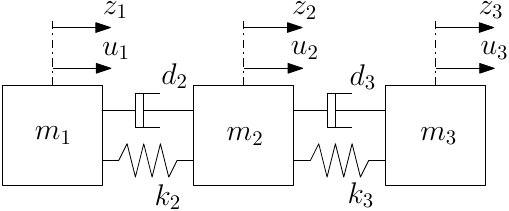}
    \caption{Simulation model of a three-mass spring damper system.}
    \label{fig: simulation model}
\end{figure}

\begin{figure}
    \centering
    \includegraphics{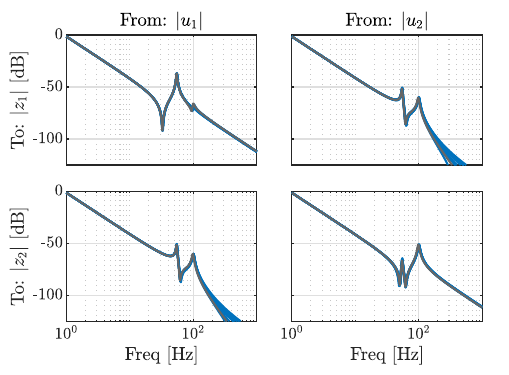}
    \vspace{-0.5em}
   \caption{Open-loop experiment: subset of the element-wise bode magnitude plot with the true system \tikzline{MatlabGray} and 10 estimation realizations~\tikzline{MatlabBlue}.}
   \label{fig: CT - experiment 1}

\includegraphics{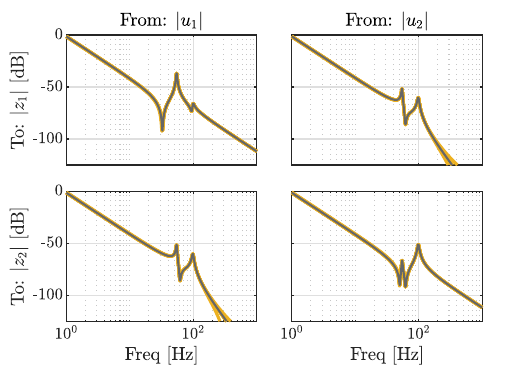}
    \vspace{-0.5em}
    \caption{Closed-loop experiment: subset of the element-wise bode magnitude plot with the true system \tikzline{MatlabGray} and 10 estimation realizations~\tikzline{MatlabYellow}.  }
    \label{fig: CT - experiment 2}
\end{figure}

This section aims to validate the introduced estimators through numerical simulations. Consider the 3-mass spring damper system in Figure \ref{fig: simulation model} with $n_{\mathrm{u}}=3$ inputs and $n_{\mathrm{y}} =3$ outputs. The model is parametrized in the additive form as
\begin{equation}
 \mathbf{G}(p, \boldsymbol{\beta}) = \frac{\mathbf{B}_{1,0}}{p^2} +  \sum^{3}_{i=2} \frac{\mathbf{B}_{i,0}}{a_{i,2}p^{2} + a_{i,1}p + 1},
\end{equation}
where the first term corresponds to a rigid-body mode, and the remaining two are flexible modes \cite{Gawronski2004AdvancedStructures}.
Two experiments are performed, with the system operating in open-loop in the first experiment and in closed-loop in the second. The external excitation signal ($\mathbf{u}(t_k)$ in the open loop and $\mathbf{r}(t_k)$ in closed loop) is a Gaussian white signal of unitary variance and the output noise $\mathbf{v}(t_k)$ is added with a variance of $0.001$ for the open-loop experiment and an output SNR of $30$ [dB] for the closed-loop experiment. For the closed-loop experiment, a low-bandwidth controller is implemented, which is discretized using a ZOH. Each parameter is initialized at a random point that deviates at most 5 percent from its true value. The model is sampled at $1$ [kHz] and in total $N=10^4$ samples are collected. The assumed intersample behavior for all signals is a ZOH and 10 realizations are computed using the estimator (27) for each setting, where each realization required approximately 2 seconds of execution time to compute. The frequency response functions (FRF) of the estimated models are shown in Figure~\ref{fig: CT - experiment 1} and Figure~\ref{fig: CT - experiment 2} for the open- and closed-loop experiments, respectively. In both cases,  the proposed estimator provides accurate estimates of the true dynamical system.

\section{Experimental validation}

\begin{figure}[t]
    \vspace{1em}
    \centering
    \includegraphics[width=0.95\columnwidth]{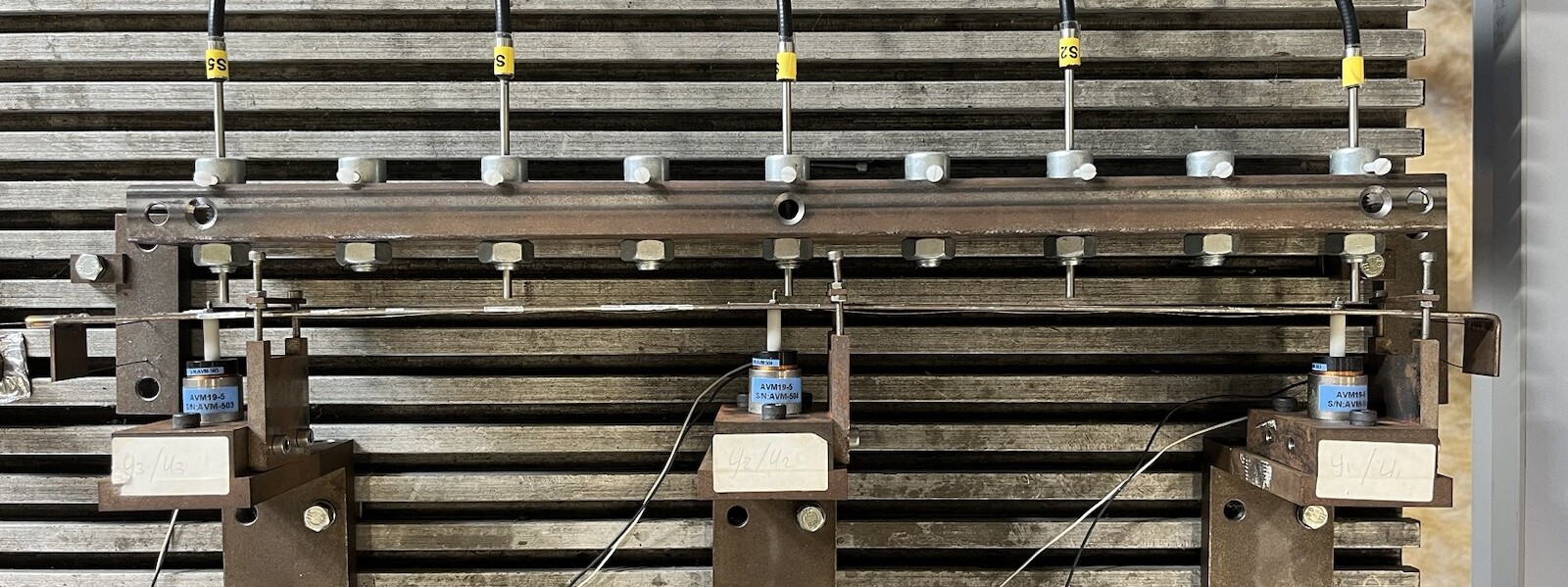}
    \vspace{1em}
    \label{fig: beam overview}
    \includegraphics[width=0.95\columnwidth]{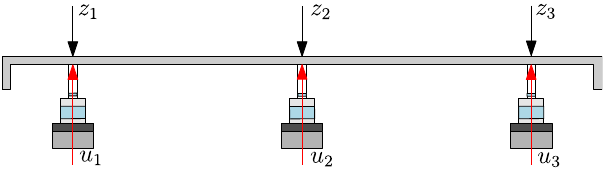}
    \vspace{-0.5em}
    \caption{The flexible beam setup (upper) and schematic overview of the featured actuators $u_i$ and sensors $z_i$ layout on the flexible beam (lower).}
    \label{fig: beam schematic}
\end{figure}

\begin{figure*}
    \centering
    \includegraphics[width = 0.95\linewidth]{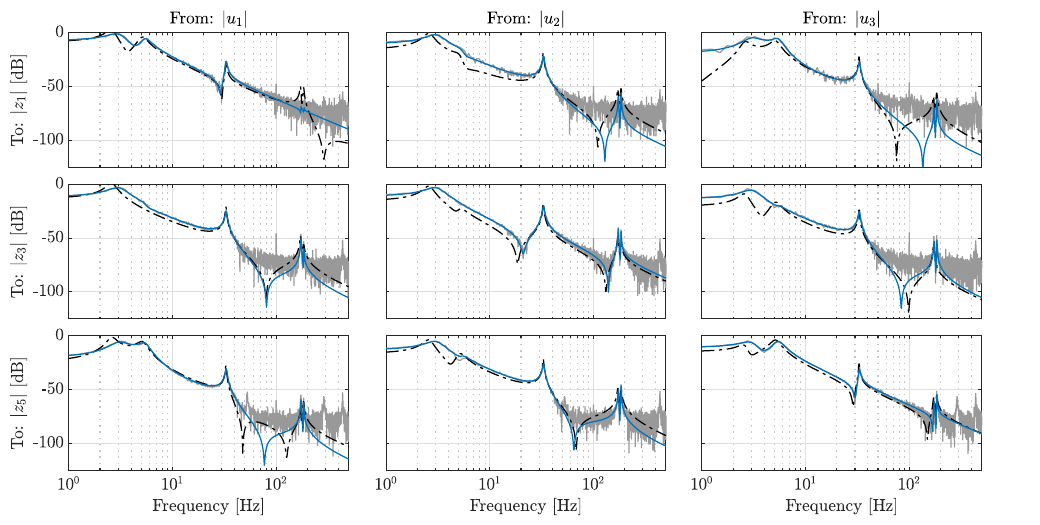}
    \vspace{-0.7em}
    \caption{Element-wise Bode magnitude plot of the the FRF measurement \tikzline{MatlabGray}, the initial estimate \tikzdashdottedline{MatlabBlack} and the estimated additive parametric model \tikzline{MatlabBlue}. The estimated model provides an accurate describtion of the flexible beam dynamics.}
    \label{fig: beam fit}
\end{figure*}
In this section, the introduced estimator is validated using experimental data, providing contribution C2. The experimental setup shown in Figure \ref{fig: beam schematic}, is a flexible beam measuring $500 \times 20 \times 2$ [mm]. The setup is actuated using three voice coil actuators and three collocated optical fiber sensors are used to measure the deflection. The system is open-loop stable, enabling the identification experiment to be conducted in an open-loop setting and operates at a sampling frequency of 4096 [Hz]. The system is excited using random phase multisine excitation signals with a flat spectrum between 0.25 [Hz] and 500 [Hz], which contain $5$ periods of $t=4$ seconds each, resulting in a total signal length of $N=81920$ samples. The assumed intersample behavior is a ZOH, and a delay of four signal samples is compensated. The model is parametrized in the additive form
\begin{equation} \label{eq: beam model}
 \mathbf{G}(p,\boldsymbol{\beta}) = \sum^{K}_{i=1} \frac{\mathbf{B}_{i,0}}{p^2/ \omega_{i}^{2} + 2(\zeta_i/\omega_i)p + 1},
\end{equation}
with $\omega_i$ the natural frequency and $\zeta_i$ the damping coefficient \cite{Gawronski2004AdvancedStructures}. The first $K=5$ modes of the system are estimated. Care must be taken when selecting \( K \), as choosing a value larger than the number of resonant modes present in the data may lead to poor convergence. The natural frequencies are initialized at the resonance peaks observed in the FRF measurement computed from the same dataset \cite{Pintelon2012SystemIdentification}, and the damping ratios are set to 1\% for each mode. The initial estimate of the numerator parameters is obtained by solving \eqref{eq: CT - initialization}, with fixed denominators based on the initial natural frequencies and damping coefficients. The FRF of the estimated additive model, which required approximately 10 seconds of execution time to compute, is shown in Figure \ref{fig: beam fit} together with the measured FRF. This comparison demonstrates that the estimated model effectively captures the true dynamics of the system, as the frequency response closely matches the measured FRF.  

A key advantage of the additive model parameterization \eqref{eq: beam model}, compared to an equivalent unfactored common denominator model of equal order, is that only second-order derivatives of the measured input/output signals are required to construct the instrument and regressor. In contrast, the common denominator model necessitates derivatives up to order ten. Additionally, the poles of the flexible beam span a wide frequency range, which causes large variations in the denominator coefficients when using an unfactored model. These aspects generally lead to more tractable computations and numerically more stable implementations.

\section{Conclusion}
This paper addresses the parametric identification of continuous-time additive MIMO systems. A unified estimator is presented, which uses a refined instrumental variable method to estimate additive models acting in either open- or closed-loop. The method enables the estimation of parsimonious and interpretable models, offering advantages when identifying physical systems. The identification strategy has been validated through numerical simulations and is successfully tested on an experimental beam setup. As directions for future work, we aim to extend the proposed method to accommodate unknown controllers, and to investigate performance under high-rate and non-uniform sampling intervals.

\bibliographystyle{IEEEtran}
\bibliography{root_arxiv.bbl}  
\end{document}